\documentclass[twocolumn,prl,showpacs]{revtex4} 
\usepackage{graphicx,amssymb,amsmath,bm}

\begin{document}

\title{Electronic Excitations in the Edge-shared Relativistic Mott 
Insulator: Na$_2$IrO$_3$}

\author {Beom Hyun Kim$^1$} 
\author {G. Khaliullin$^2$}
\author {B. I. Min$^1$}
\affiliation{$^1$Department of Physics, PCTP, 
Pohang University of Science and Technology, Pohang 790-784, Korea\\
$^2$Max Planck Institute for Solid State Research, 
Heisenbergstrasse 1, D-70569 Stuttgart, Germany}
\date{\today}

\begin{abstract}
We have investigated the excitation spectra of $j_{eff}$=$\frac{1}{2}$ Mott 
insulator Na$_2$IrO$_3$. Taking into account a relativistic multiplet 
structure of Ir ions, we have calculated the optical conductivity 
$\sigma(\omega)$ and resonant inelastic x-ray scattering (RIXS) spectra,
which manifest different features from those of a canonical 
$j_{eff}$=$\frac{1}{2}$ system Sr$_2$IrO$_4$. Distinctly from the 
two-peak structure in Sr$_2$IrO$_4$, $\sigma(\omega)$ in Na$_2$IrO$_3$ has 
a broad single peak dominated by interband transitions from
$j_{eff}$=$\frac{3}{2}$ to $\frac{1}{2}$. 
RIXS spectra exhibit the spin-orbit (SO) exciton
that has a two-peak structure arising from the crystal-field effect, 
and the magnon peak at energies much lower than in Sr$_2$IrO$_4$.
In addition, a small peak near the optical absorption edge is found in 
RIXS spectra, originating from the coupling between the electron-hole 
($e$-$h$) excitation and the SO exciton. Our findings corroborate the 
validity of the relativistic electronic structure and importance of both 
itinerant and local features in Na$_2$IrO$_3$.
\end{abstract}


\maketitle

\begin{figure}[b]
\centering
\includegraphics[width=6.8 cm]{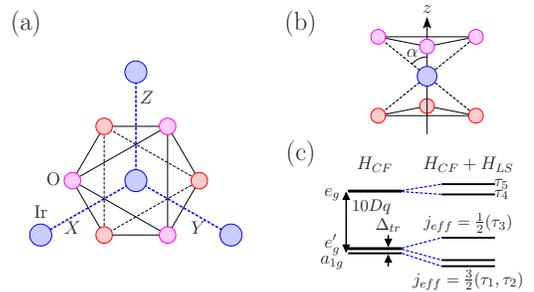}
\caption
{ (Color online)
   (a)  Top view of edge-shared Na$_2$IrO$_3$.
   $X$, $Y$, and $Z$ represent directions of nearest neighboring Ir sites.
   (b) The trigonally compressed IrO$_6$ octahedron, where the angle 
   $\alpha$ between $z$-axis and Ir-O bond direction is about $57.96^{\circ}$,  
   instead of $\cos^{-1}\sqrt{\frac{1}{3}}\approx54.74^{\circ}$ for
   $O_h$ symmetry. 
   (c) Energy splitting of local $d$ levels in the presence of 
   the trigonal distortion and SO coupling.
}
\label{fig1}
\end{figure}

Rich physical properties in $4d$ and $5d$ transition metal (TM) oxides 
arise from the mutual interplay of electronic degrees of freedom
such as bandwidth $W$, Coulomb correlation $U$,
and spin-orbit (SO) coupling $\lambda$~\cite{Pesin}.
Sr$_2$IrO$_4$ is one of the most-studied $5d$ TM systems to examine 
cooperative effects of the electronic degrees of freedom, 
which yield the intriguing $j_{eff}$=$\frac{1}{2}$ Mott insulating 
nature~\cite{BJKim1,BJKim2,Jackeli,JKim,BHKim}.
Another iridate Na$_2$IrO$_3$ also draws the recent attention
because of its insulating nature similar to that of Sr$_2$IrO$_4$.
In contrast to Sr$_2$IrO$_4$ with corner-shared IrO$_6$ octahedra,
Na$_2$IrO$_3$ is composed of edge-shared octahedra (see Fig.~\ref{fig1}), 
and Ir ions form a honeycomb lattice. Early proposals that Na$_2$IrO$_3$ 
may be a topological insulator~\cite{Shitade} or host Kitaev model
physics~\cite{Jackeli} triggered a theoretical and experimental activity aimed
to understand insulating nature and magnetic structure of 
Na$_2$IrO$_3$~\cite{Jin,Singh1,Chaloupka1,Liu,Singh2,Choi,Ye,Kimchi,Chaloupka2,
Comin,Mazin1}.

The strong SO coupling in iridates causes $t_{2g}$ orbitals to 
split into $j_{eff}$=$\frac{1}{2}$ and $\frac{3}{2}$ states 
[see Fig.~\ref{fig1}(c)],
and then the resulting narrow half-filled $j_{eff}$=$\frac{1}{2}$ band is to be 
split even by a weak Coulomb repulsion to become a Mott insulator~\cite{BJKim1}.
This scenario, however, has been questioned recently by 
Mazin \textit{et al.}\cite{Mazin1,Mazin2,Foyevtsova},
who argued that the insulating nature of Na$_2$IrO$_3$ originates from
the formation of quasi-molecular orbital (QMO) states of Ir hexagon,
which have a considerable itinerant character. Physically, this kind of 
controversy is evoked due to dual (atomic/band) nature of
Ir $5d$ orbitals. Because three relevant physical parameters 
$W$, $\lambda$, and $U$ of Ir $5d$ orbitals are of similar energy scale, 
it is not easy to identify which parameter is dominant in determining 
the electronic structures of iridates.

In fact, dual nature of $5d$ orbitals is reflected on excitations, which 
manifest various peculiar features in iridates. In the case of Sr$_2$IrO$_4$, 
the local $d$-$d$ transition between $j_{eff}$=$\frac{1}{2}$ and 
$\frac{3}{2}$, termed as the SO exciton, is observed in resonant 
inelastic x-ray scattering (RIXS) spectra~\cite{JKim},  and optical 
conductivity $\sigma(\omega)$ exhibits a prominent two-peak structure 
at 0.5 and 1.0 eV in the vicinity of Mott gap region~\cite{Moon}.
$\sigma(\omega)$ and RIXS spectra for Na$_2$IrO$_3$ display some features 
distinct from those in Sr$_2$IrO$_4$. Instead of a two-peak structure, 
$\sigma(\omega)$ of Na$_2$IrO$_3$ shows just a broad single peak at higher 
energy of $\sim$ 1.5 eV~\cite{Comin,Sohn}. SO exciton peak is present in RIXS 
spectra of Na$_2$IrO$_3$ too, but it has a well resolved two-peak structure 
and a negligible momentum dependence. The origin of these two RIXS-peaks 
is under debate, whether they come
from the trigonal crystal-field~\cite{Gretarsson} or not~\cite{Foyevtsova}.
In addition, an extra RIXS-peak was detected in Na$_2$IrO$_3$ near the optical
absorption edge ($\sim$ 0.4 eV), whose origin is not yet settled.  
It is important to examine the similarities and differences 
between Sr$_2$IrO$_4$ and Na$_2$IrO$_3$ including both correlation and 
itineracy effects, in order to clarify what kind of electronic nature 
prevails in these compounds: atomic, band, or dual nature.

In this Letter, we have investigated characteristic features of excitation 
spectra in Na$_2$IrO$_3$. More specifically, we addressed the following 
questions currently in dispute: (i) why $\sigma(\omega)$ has a single-peak 
structure distinctly from that of Sr$_2$IrO$_4$, 
(ii) what is the origin of two peaks of the SO exciton in RIXS spectra,
and (iii) what is the identity of an extra RIXS peak observed near the 
optical edge. For this purpose, we have generated the microscopic model 
incorporating the full local multiplets of Ir ions and their hopping integrals.
Using the exact diagonalization (ED) method, we have calculated 
$\sigma(\omega)$ and RIXS spectra for Na$_2$IrO$_3$, and extracted 
the physical parameters that best describe the experimental data.
We have demonstrated that the coupling between the itinerant $e$-$h$
excitations and the local SO exciton is essential in Na$_2$IrO$_3$ but,  
due to different hopping topology in a honeycomb lattice, the 
manifestations of this effect in $\sigma(\omega)$ and RIXS spectra are 
different from those in Sr$_2$IrO$_4$. 


\begin{table}[b]
\caption
{Physical parameters of Na$_2$IrO$_3$ in units of eV. They are adopted 
to be consistent with literature ($\Delta$~\cite{Comin}, 
$\Delta_{tr}$~\cite{Foyevtsova}) and to optimize theoretical RIXS spectra 
($10Dq$, $J_H$, $\lambda$, $t_{pd\sigma}$) and $\sigma(\omega)$ ($U$).
} \label{Para}
\begin{ruledtabular}
\begin{tabular}{c c c c c c c c }
 $10Dq$ & $\Delta$ & $\Delta_{tr}$ & $U$ & $J_H$ & $\lambda$ & 
 $t_{pd\sigma}$ & $t_{pd\pi}$ \\
\hline
3.3 & 3.3 & 0.075 & 1.35 & 0.25 & 0.43 & $-$1.90& $-0.46t_{pd\sigma}$ \\
\end{tabular}
\end{ruledtabular}
\end{table}

\begin{figure}[t]
\centering
\includegraphics[width=7.6 cm]{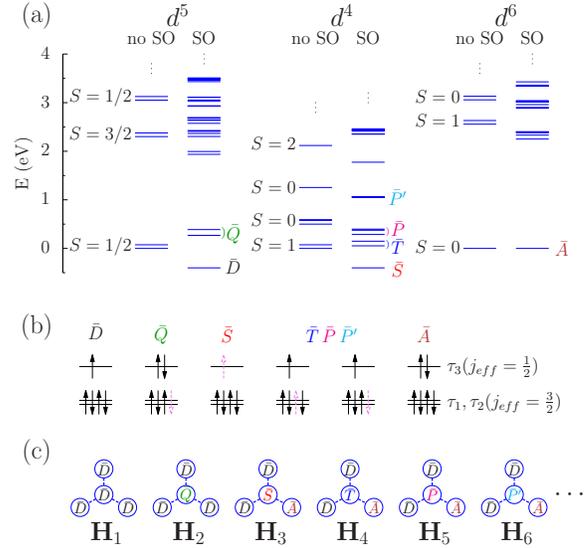}
\caption
{ (Color online) 
  (a) Relative energies for $d^4$, $d^5$, and $d^6$ multiplets
  of Na$_2$IrO$_3$ with physical parameters from Table I.
  When the trigonal distortion is absent ($\Delta_{tr}$=0.0),
  each $\bar{Q}$, $\bar{T}$, $\bar{P}$, and $\bar{P}^{\prime}$ multiplets
  are degenerate.
  (b) Relevant configurations that give dominant contributions to 
  low energy multiplets.  
  Violet dotted arrows represent removed spins from $\bar D$ multiplet.
  Because $10Dq$ is large enough, 
  isospins in relevant multiplets occupy three double group levels 
  ($\tau_1,\tau_2,\tau_3$), which are mainly attributed to
  $t_{2g}$ manifolds ($j_{eff}$=$\frac{1}{2}$, $j_{eff}$=$\frac{3}{2}$).
  But we consider all the five double group states ($\tau_1$-$\tau_5$) 
  in describing the multiplets.
  (c) Schematic diagrams of possible cluster
  multiplets included in each subspace.
}
\label{fig2}
\end{figure}

To investigate electronic structures of two dimensional honeycomb lattice
Na$_2$IrO$_3$, we considered a four-site Ir cluster as shown 
in Fig.~\ref{fig1}(a). The local Hamiltonian of a Ir site reads as:
\begin{align}
\label{eq1}
&H_{ion} =\sum_{\mu\sigma} \epsilon_{\mu} n_{\mu\sigma}+
   \lambda \sum_{\mu\nu\sigma\sigma^{\prime}}
   (\mathbf{l}\cdot\mathbf{s})_{\mu\sigma,\nu\sigma^{\prime}} 
   c_{\mu\sigma}^{\dagger}c_{\nu\sigma^{\prime}} \nonumber  \\
  &+ \frac{1}{2}\sum_{\sigma\sigma^{\prime}\mu\nu} 
  U_{\mu\nu} c_{\mu\sigma}^{\dagger}c_{\nu\sigma^{\prime}}^{\dagger}
  c_{\nu\sigma^{\prime}}c_{\mu\sigma}  
  + \frac{1}{2}\sum_{\substack{\sigma \sigma^{\prime} \\ \mu\ne\nu}}
  J_{\mu\nu}c_{\mu\sigma}^{\dagger}c_{\nu\sigma^{\prime}}^{\dagger}
  c_{\mu\sigma^{\prime}}c_{\nu\sigma} \nonumber \\
  & + \frac{1}{2}\sum_{\substack{\sigma \\ \mu\ne\nu}} 
  J_{\mu\nu}^{\prime} c_{\mu\sigma}^{\dagger}c_{\mu-\sigma}^{\dagger}
  c_{\nu-\sigma}c_{\nu\sigma},  
\end{align}
where $\mu$ and $\sigma$ refer to orbital and spin states of Ir, respectively.
Because of the trigonal distortion~\cite{Octa} and the strong SO coupling
(first and second terms in Eq.~\ref{eq1}),
Ir 5$d$ orbitals are split into five double group states 
($\tau_1$-$\tau_5$), as shown in Fig.~\ref{fig1}(c)~\cite{DG}.
$U_{\mu\nu}$ , $J_{\mu\nu}$ , and $J_{\mu\nu}^{\prime}$ are 
direct Coulomb, exchange Coulomb, and pair hopping integrals,
which can be given by $U$ and $J_H$ parameters~\cite{Hmail}.
Figure~\ref{fig2}(a) presents local electronic energies of Ir multiplets
calculated with physical parameters in Table~\ref{Para}. 
Because $10Dq$ is large enough ($\sim3.3$ eV), 
the lowest three double group states ($\tau_1$,$\tau_2$,$\tau_3$)
mainly contribute to low energy multiplets of $d^4$, $d^5$, and $d^6$
configurations, see Fig.~\ref{fig2}(b). 
In describing the Hilbert space of the four-site cluster,
we took into account several lowest multiplets, \textit{e.g.},
six ($\bar{D}$,$\bar{Q}$) for $d^5$, 
fourteen ($\bar{S}$,$\bar{T}$,$\bar{P}$,$\bar{P}^\prime$) 
for $d^4$, and one ($\bar{A}$) for $d^6$.
Note that $\bar{S}$ corresponds to the hole state mainly of 
$j_{eff}$=$\frac{1}{2}$ band,
whereas $\bar{T}$, $\bar{P}$, and $\bar{P}^{\prime}$ correspond to those of
$j_{eff}$=$\frac{3}{2}$ bands.

In order to save computational cost, we restricted the Hilbert space into 
all possible multiplets of $d^5$-$d^5$-$d^5$-$d^5$ and 
$d^4$-$d^6$-$d^5$-$d^5$ configurations, 
which are expressed as the direct product of relevant multiplets of 
Ir ions ($\bar{D}\cdots\bar{A}$). 
Because this restricted space already 
includes all possible states with energies lower than $2.0$ eV, it will provide
appropriate details of low energy excitations in Na$_2$IrO$_3$. To discern 
excitation distributions, we classified the Hilbert space into seven subspaces:
$\mathbf{H}_1$-$\mathbf{H}_7$~\cite{Hspace}.
Some examples included in each subspace are shown in Fig.~\ref{fig2}(c).
To include the itineracy effects, we considered the hopping between nearest 
neighboring (NN) Ir's via intermediate oxygen~\cite{Hop}. 
Employing the Slater-Koster theory~\cite{Slater},
we calculated the $pd$-hopping matrix in terms of
$t_{pd\sigma}$ and $t_{pd\pi}$ parameters and evaluated the effective hopping
$t_{dd}(\tau\tilde{\sigma};\tau^{\prime}\tilde{\sigma}^{\prime})$
between NN double group states $\tau\tilde{\sigma}$
and $\tau^{\prime}\tilde{\sigma}^{\prime}$ by summing 
$\sum_{p\sigma}\frac{t_{pd}(\tau\tilde{\sigma};p\sigma)
 t_{pd}^*(\tau^{\prime}\tilde{\sigma}^{\prime};p\sigma)}
{\sqrt{(\Delta+\epsilon_{\tau})(\Delta+\epsilon_{\tau^{\prime}})}}$ values of 
two Ir-O-Ir paths ($\Delta$ is $p$-$d$ charge transfer energy)~\cite{Hop2}.

Using the ED method, we have solved the Hamiltonian of the four-site 
cluster and investigated the excitation spectra. Let $E_n$ and 
$|\Psi_n\rangle$ be the $n$-th eigenvalue and the eigenvector of the cluster, 
respectively. To examine the excitation distribution, we obtained the 
projected excitation spectrum (PES) 
as $\Lambda_i(\omega) = \sum_n \sum_{m\in \mathbf{H}_i}
|\langle \Psi_n | m\rangle|^2 \delta(\omega-E_n)$, where
$|m\rangle$ represents the orthonormal basis of the subspace
$\mathbf{H}_i$. To compare theoretical PES to observed excitations in 
Na$_2$IrO$_3$, we calculated $\sigma\left(\omega\right)$ and RIXS spectra 
by using the Kubo formula~\cite{BHKim}. We set $k_B T=30$ meV.

\begin{figure}[!t]
\centering
\includegraphics[width=7.0 cm]{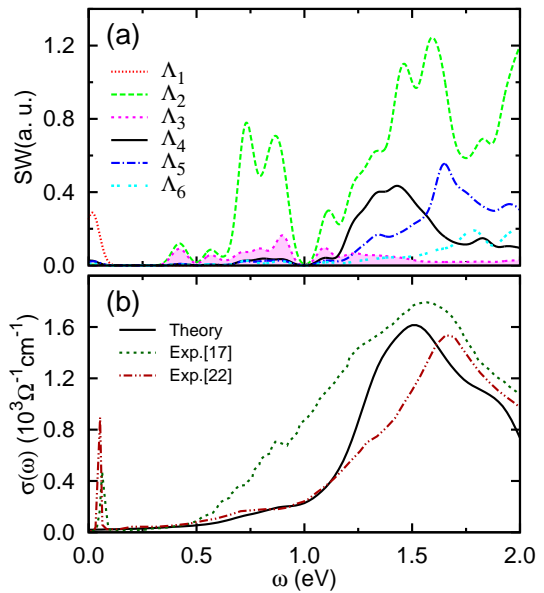}
\caption
{ (Color online)
   (a) Spectral weight (SW) of projected excitation spectra (PES) for 
   Na$_2$IrO$_3$.
   (b) Optical conductivity of Na$_2$IrO$_3$ as calculated (solid line) and
   measured at $T=300$~K (dotted line~\cite{Comin}, dot-dashed 
   line~\cite{Sohn}). The observed sharp peak near 0.1 eV is of
   phonon origin (not included in our calculations).
}
\label{fig3}
\end{figure}

Figure~\ref{fig3}(a) shows the PES for Na$_2$IrO$_3$. Let us first explore 
the relation between a specific PES and each excitation. Because 
$\mathbf{H}_1$ includes all possible fluctuations of isospin $J_{eff}$=$1/2$ 
for $d^5$, $\Lambda_1$ ($\bar{D}\bar{D}\bar{D}\bar{D}$) represents the 
magnon excitation. $\Lambda_2$ ($\bar{D}\bar{Q}\bar{D}\bar{D}$) represents 
one or more SO excitons because $\bar{Q}$ is one hole state of 
$j_{eff}$=$\frac{3}{2}$ for $d^5$ [Fig.~\ref{fig2}(b)].
$\Lambda_3$ ($\bar{A}\bar{S}\bar{D}\bar{D}$) and $\Lambda_4$-$\Lambda_6$ are 
related to the $e$-$h$ excitations involving hole states of 
$j_{eff}$=$\frac{1}{2}$ ($\bar{S}$) and $j_{eff}$=$\frac{3}{2}$ 
($\bar{T}$, $\bar{P}$, $\bar{P}^{\prime}$), respectively.

\begin{figure}[!t]
\centering
\includegraphics[width=7.0 cm]{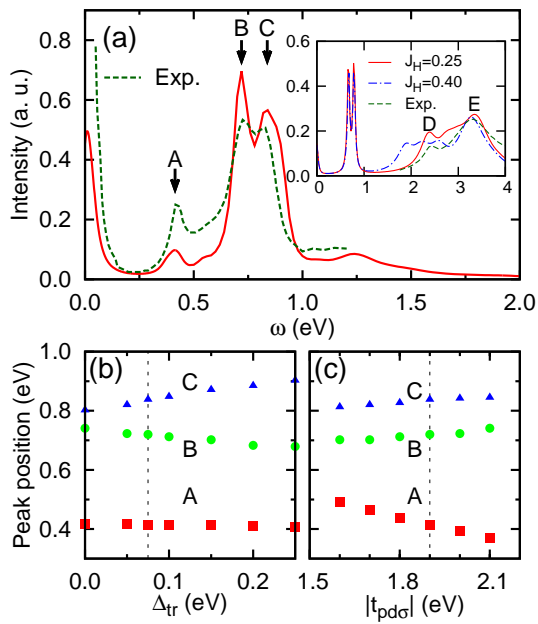}
\caption
{ (Color online)
   (a) Calculated RIXS spectra (solid line) for the four-site cluster of 
   Na$_2$IrO$_3$. Dashed line: experimental data measured at room 
   temperature~\cite{Gretarsson}. 
   Inset: calculated RIXS spectra for a single-site cluster 
   incorporating all possible $d^5$ multiplets at two different $J_H$.
   Peak positions as functions of (b) the trigonal distortion and (c) 
   the hopping strength.
   Vertical lines in (b) and (c) denote the possible optimal parameters
    of $\Delta_{tr}$ and $t_{pd\sigma}$.
}
\label{fig4}
\end{figure}

We can notice interesting features in the PES of Fig.~\ref{fig3}.
(1) $\Lambda_1$ shows a magnon peak at low energies, below $50$ meV.
This feature is very different from that of Sr$_2$IrO$_4$, in which 
magnon spectra spread over $0$-$250$ meV~\cite{BHKim}.
It implies strong suppression of the magnetic interaction in Na$_2$IrO$_3$
due to its edge-shared bonding nature. Because the Ir-O-Ir bond angle is 
nearly 90$^{\circ}$, the effective hopping between 
$j_{eff}$=$\frac{1}{2}$ ($\tau_3$) states is suppressed a lot. Actually, 
this hopping is almost one-order of magnitude smaller than that between 
$j_{eff}$=$\frac{1}{2}$ and $\frac{3}{2}$ ($\tau_1$) states~\cite{Hop2}.
(2) $\Lambda_2$ exhibits two peaks at 0.73 and 0.86 eV.
This spectrum is attributed to the on-site $d$-$d$ transition from
occupied $j_{eff}$=$\frac{3}{2}$ ($\tau_1$,$\tau_2$) to unoccupied 
$j_{eff}$=$\frac{1}{2}$ ($\tau_3$), which is reminiscent of the SO exciton 
in Sr$_2$IrO$_4$. Indeed, as will be shown in Fig.~\ref{fig4}(a), this 
spectrum is consistent with experimental RIXS peak positions for 
Na$_2$IrO$_3$~\cite{Gretarsson}. $\Lambda_2$ has peaks above $1.0$ eV too.
They, however, hardly produce RIXS spectra because
they correspond to two or more simultaneous SO excitons.
(3) $\Lambda_3$ spreads over broad energy range above $\omega\approx0.4$ eV.
It does not look like a single peak corresponding to
the $\bar{A}\bar{S}\bar{D}\bar{D}$ multiplet,
which indicates that simple atomic picture is inadequate 
to describe the $j_{eff}$=$\frac{1}{2}$ $e$-$h$ excitation of Na$_2$IrO$_3$. 
There should be considerable mixing among a few multiplets due to itinerant 
character of Ir $5d$ bands. Moreover, $\Lambda_3$ shows a small peak near 
the $e$-$h$ excitation edge ($\omega\sim0.4$ eV). Interestingly, 
$\Lambda_2$ also has a peak in the same region with almost the same intensity.
This feature suggests that there is a strong mixing between $\Lambda_2$ and 
$\Lambda_3$, which is supposed to produce both the broad dispersion 
and the edge state in $\Lambda_3$.
(4) $\Lambda_4$-$\Lambda_6$ are distributed above $\omega=$1.2 eV.
Despite their broad dispersions, each PES has its own predominant peak,
implying that local multiplets of $j_{eff}$=$\frac{3}{2}$ hole are retained.
As shown in Fig.~\ref{fig3}(b), in this region of $\sim1.5$ eV, 
there appears a broad peak of $\sigma(\omega)$ in Na$_2$IrO$_3$.
It is thus expected that the interband $e$-$h$ transitions 
from $j_{eff}$=$\frac{3}{2}$ to $\frac{1}{2}$ 
($\bar{T}$,$\bar{P}$,$\bar{P}^{\prime}$) give rise to main
spectral weight in $\sigma(\omega)$ of Na$_2$IrO$_3$.

Figure~\ref{fig3}(b) presents theoretical result for 
$\sigma\left(\omega\right)$. Similar to experimental data, it exhibits 
a predominant peak at around 1.5 eV, which certainly reflects the major role
of $j_{eff}$=$\frac{3}{2}$ states, as explained above. The spectral weight of 
$j_{eff}$=$\frac{1}{2}$ band (which dominates in 
Sr$_2$IrO$_4$) is suppressed because of hopping topology of edge-shared 
Na$_2$IrO$_3$. This behavior in Na$_2$IrO$_3$ is contrary to that in 
Sr$_2$IrO$_4$, for which two prominent peaks of $j_{eff}$=$\frac{1}{2}$ band  
origin appear through the Fano-type overlap between the $e$-$h$ continuum 
of the $j_{eff}$=$\frac{1}{2}$ band and the on-site SO exciton~\cite{BHKim}.

Shown in Figure~\ref{fig4}(a) is the theoretical RIXS spectra at 
$\mathbf{q}=0$~\cite{Rixs}. Noteworthy is the emergence of three-peak 
structure (denoted by $A$, $B$, and $C$), which is consistent with the 
experiment. To elucidate the origin of these three peaks, 
we also calculated RIXS spectra for a single-site IrO$_6$ cluster, including
all possible $d^5$ multiplets. In this case, only two peaks appear at 0.67 and 
0.78 eV [see inset of Fig.~\ref{fig4}(a)], which are equivalent 
to $B$ and $C$ peaks for the four-site cluster. This implies that both 
$B$ and $C$ correspond to local excitations, which are attributed to
on-site $d$-$d$ transitions between $j_{eff}=\frac{1}{2}$ and
$\frac{3}{2}$ orbitals. Then it is natural to conjecture that the 
energy difference between $B$ and $C$ comes from the crystal-field splitting 
of $j_{eff}$=$\frac{3}{2}$ states. Indeed, as shown in Fig.~\ref{fig4}(b),
the splitting between $B$ and $C$ increases with increasing the trigonal 
distortion strength of $\Delta_{tr}$. We point out that the observed $B$-$C$ 
splitting of the order of 110 meV is well explained by our calculations which
include correlation effects, even though we used a rather small input value of 
$\Delta_{tr}=75$~meV~\cite{Foyevtsova}. This can be understood as a 
correlation-induced enhancement of the crystal-field splitting~\cite{Pot07}. 

It is seen in Fig.~\ref{fig4}(a) that the peak $A$ near 0.4 eV is missing 
in the single-site calculation of inset. This finding suggests that the 
peak $A$ is related to itinerant nature of Ir $5d$ orbitals, especially, 
the inter-site hopping between $j_{eff}=\frac{1}{2}$ and $\frac{3}{2}$ states,
which is dominant here~\cite{Hop2}. We already noticed in the PES of 
Fig.~\ref{fig3}(a) that there is a strong coupling between SO exciton 
$\Lambda_2$ and $e$-$h$ continuum $\Lambda_3$ in the vicinity of peak $A$.
More convincing evidence is found in Fig.~\ref{fig4}(c), which presents the 
peak positions as a function of the hopping strength. We note that the larger 
the hopping strength is, the lower the peak position of $A$ is.
This behavior reveals that the peak $A$ at the edge of $e$-$h$ excitation 
certainly originates from the inter-site hopping, which brings about the 
coupling of broad $e$-$h$ continuum with the local SO exciton.
Note that, this $e$-$h$ excitation is hardly detectable in the optical spectra. 
It is due to the large suppression of the direct hopping between 
$j_{eff}=\frac{1}{2}$ bands in the edge-shared of Na$_2$IrO$_4$.
This finding is different from a suggestion of Ref.~\onlinecite{Gretarsson} 
that the peak $A$ is an excitonic bound state due to long-range Coulomb 
interaction.

The single-site calculation in the inset of Fig.~\ref{fig4}(a) also gives 
$D$ and $E$ peaks above 2.0 eV, which are in good agreement with 
experiment~\cite{Gretarsson}. Note that the peak position of $D$ moves 
with varying $J_H$, while that of $E$ does not. Both $D$ and $E$ correspond 
to local excitations to $t_{2g}^4e_g^1$ configurations. But they have 
different spin states: $D$ has high-spin $S$=$\frac{3}{2}$ while  
$E$ has low-spin $S$=$\frac{1}{2}$, as shown in Fig.~\ref{fig2}(a) for $d^5$.
Energies of the former and the latter with respect to the ground state
are given by $10Dq-4J_H$ and $10Dq$, respectively. Accordingly, from the peak 
positions of $D$ and $E$, one can determine $10Dq$ and $J_H$ values. 

Our RIXS calculation yields the magnetic peak below $\sim$50 meV, in a
qualitative agreement with the recent RIXS experiment for Na$_2$IrO$_3$,
where the magnetic excitations dispersing up to energy of $\sim$35 meV have been 
observed~\cite{Gretarsson2}. The agreement is not surprising since our 
calculations fully include the exchange processes discussed in 
Ref.~\onlinecite{Chaloupka2} that contribute to Kitaev-Heisenberg 
interactions (except a direct hopping between Ir's which is 
small~\cite{Mazin1}).  

Our successful description of various excitations in Na$_2$IrO$_3$ indicate 
the realization of the relativistic electronic structure in this material.
According to recent DFT calculation~\cite{HSKim}, the majority of Wannier 
orbitals near the Fermi level have indeed a dominant $j_{eff}$=$\frac{1}{2}$ 
character, with only small $j_{eff}$=$\frac{3}{2}$ tails on the NN sites. 
More importantly, the results presented above make it clear that both 
itinerant and local features have to be accounted for to describe the 
experimental observations.  


In conclusion, we have clarified controversial issues of Na$_2$IrO$_3$,
by unraveling the identities of low energy excitations observed in 
$\sigma\left(\omega\right)$ and RIXS spectra. The broad peak of 
$\sigma\left(\omega\right)$ in Na$_2$IrO$_3$ is attributed to $e$-$h$ 
excitations from $j_{eff}$=$\frac{3}{2}$ to $\frac{1}{2}$ bands, 
in contrast to Sr$_2$IrO$_4$ where two-peak structure arises from $e$-$h$ 
excitations of $j_{eff}$=$\frac{1}{2}$ band through the Fano-type overlap 
with the on-site SO exciton. Two peaks at 0.7-0.8 eV in RIXS spectra of 
Na$_2$IrO$_3$ come from local $d$-$d$ transitions between two relativistic 
states, and their splitting is caused by the trigonal crystal-field 
enhanced by correlation effects. The RIXS peak found in Na$_2$IrO$_3$ 
near $\omega\sim0.4$~eV originates from the coupling between
the $e$-$h$ excitation of $j_{eff}$=$\frac{1}{2}$ band and the SO exciton
in the vicinity of optical absorption edge. Altogether, our study confirms the 
relativistic Mott insulating nature of Na$_2$IrO$_3$, and demonstrates that 
the Fano-type coupling between the itinerant $e$-$h$ excitations and 
the local SO transition is an intrinsic nature in iridate systems including 
both Na$_2$IrO$_3$ and Sr$_2$IrO$_4$.

We thank B. J. Kim and Heung-Sik Kim for fruitful discussions.
This work was supported by the NRF (No.2009-0079947).

\newpage

\renewcommand{\thefigure}{S\arabic{figure}}
\renewcommand{\thetable}{S\arabic{table}}
\setcounter{table}{0}

\onecolumngrid

\begin{center}
{\bf \Large
{\it Supplemental Material:}\\
Electronic excitations in the edge-shared relativistic Mott insulator: 
Na$_2$IrO$_3$
}
\end{center}

\section{Relativistic orbital states}

In the trigonal distortion (see Fig.~1(b)), 
$d$ orbital states are split into the following eigenstates:
\begin{align}
|e_{g1}^{\prime}\rangle&= \cos\alpha^{\prime}|xy\rangle
         -\sin\alpha^{\prime}|yz\rangle, \nonumber \\
|e_{g2}^{\prime}\rangle&= \cos\alpha^{\prime}|x^2-y^2\rangle
         +\sin\alpha^{\prime}|zx\rangle, \nonumber \\
|a_{1g}\rangle &= |z^2\rangle, \nonumber  \\
|e_{g1}\rangle&= \cos\alpha^{\prime}|zx\rangle
         -\sin\alpha^{\prime}|x^2-y^2\rangle, \nonumber \\
|e_{g2}\rangle &= \cos\alpha^{\prime}|yz\rangle
         +\sin\alpha^{\prime}|xy\rangle. \nonumber 
\end{align}
Note that, when $\cos\alpha^{\prime}=\sqrt{\frac{2}{3}}\approx0.816$,
$e_{g}^{\prime}$ and $a_{1g}$ orbitals correspond to 
$t_{2g}$ orbitals in the local $O_h$ symmetry.
In the presence of the trigonal distortion,
$\cos\alpha^{\prime}$ is to be reduced from $\sqrt{\frac{2}{3}}$.
In our calculation for Na$_2$IrO$_3$,
we have adopted $\cos\alpha^{\prime}\approx0.776$
to fit $10Dq$ and $\Delta_{tr}$ parameters based on 
the crystal field calculation\cite{Yosida}. 
Thus, the double group states for given parameters in Table~I are expressed 
as following:
\begin{align}
|\tau_{1}\tilde{\uparrow}\rangle &\approx0.811|a_{1g}\uparrow\rangle-
  0.402i\left(|e_{g1}^{\prime}\downarrow\rangle+
    i|e_{g2}^{\prime}\downarrow\rangle\right)+
  0.098\left(|e_{g1}\downarrow\rangle+i|e_{g2}\downarrow\rangle\right),
  \nonumber \\
|\tau_2\tilde{\uparrow}\rangle &\approx 
  0.437\left(|e_{g1}^{\prime}\uparrow\rangle+
    i|e_{g2}^{\prime}\uparrow\rangle\right)+
  0.028i\left(|e_{g1}\uparrow\rangle+i|e_{g2}\uparrow\rangle\right)-
  0.545\left(|e_{g1}^{\prime}\downarrow\rangle-
    i|e_{g2}^{\prime}\downarrow\rangle\right)-
  0.103i\left(|e_{g1}\downarrow\rangle-i|e_{g2}\downarrow\rangle\right),
 \nonumber \\
|\tau_{3}\tilde{\uparrow}\rangle &\approx
  0.579\left(|e_{g1}^{\prime}\uparrow\rangle-
    i|e_{g2}^{\prime}\uparrow\rangle\right)-
  0.006i\left(|e_{g1}\uparrow\rangle-i|e_{g2}\uparrow\rangle\right)-
  0.575i|a_{1g}\downarrow\rangle,
 \nonumber \\
|\tau_4\tilde{\uparrow}\rangle &\approx 
  -0.077\left(|e_{g1}^{\prime}\uparrow\rangle+
    i|e_{g2}^{\prime}\uparrow\rangle\right)-
  0.690i\left(|e_{g1}\uparrow\rangle+i|e_{g2}\uparrow\rangle\right)-
  0.075\left(|e_{g1}^{\prime}\downarrow\rangle-
    i|e_{g2}^{\prime}\downarrow\rangle\right)-
  0.115i\left(|e_{g1}\downarrow\rangle-i|e_{g2}\downarrow\rangle\right),
 \nonumber \\
|\tau_{5}\tilde{\uparrow}\rangle &\approx
  0.061\left(|e_{g1}^{\prime}\uparrow\rangle-
    i|e_{g2}^{\prime}\uparrow\rangle\right)+
  0.700i\left(|e_{g1}\uparrow\rangle-i|e_{g2}\uparrow\rangle\right)+
  0.109i|a_{1g}\downarrow\rangle,
  \nonumber \\
|\tau_{1}\tilde{\downarrow}\rangle &\approx
  0.402\left(|e_{g1}^{\prime}\uparrow\rangle-
    i|e_{g2}^{\prime}\uparrow\rangle\right)-
  0.098i\left(|e_{g1}\uparrow\rangle-i|e_{g2}\uparrow\rangle\right)+
  0.811i|a_{1g}\downarrow\rangle,
 \nonumber \\
|\tau_2\tilde{\downarrow}\rangle &\approx 
  0.545\left(|e_{g1}^{\prime}\uparrow\rangle+
    i|e_{g2}^{\prime}\uparrow\rangle\right)-
  0.103i\left(|e_{g1}\uparrow\rangle+i|e_{g2}\uparrow\rangle\right)+
  0.437\left(|e_{g1}^{\prime}\downarrow\rangle-
    i|e_{g2}^{\prime}\downarrow\rangle\right)-
  0.028i\left(|e_{g1}\downarrow\rangle-i|e_{g2}\downarrow\rangle\right),
 \nonumber \\
|\tau_3\tilde{\downarrow}\rangle &\approx0.575|a_{1g}\uparrow\rangle+
  0.579i\left(|e_{g1}^{\prime}\downarrow\rangle+
    i|e_{g2}^{\prime}\downarrow\rangle\right)-
  0.006\left(|e_{g1}\downarrow\rangle+i|e_{g2}\downarrow\rangle\right),
 \nonumber \\
|\tau_4\tilde{\downarrow}\rangle &\approx 
  0.075\left(|e_{g1}^{\prime}\uparrow\rangle+
    i|e_{g2}^{\prime}\uparrow\rangle\right)-
  0.115i\left(|e_{g1}\uparrow\rangle+i|e_{g2}\uparrow\rangle\right)-
  0.077\left(|e_{g1}^{\prime}\downarrow\rangle-
    i|e_{g2}^{\prime}\downarrow\rangle\right)+
  0.690i\left(|e_{g1}\downarrow\rangle-i|e_{g2}\downarrow\rangle\right),
 \nonumber \\
|\tau_5\tilde{\downarrow}\rangle &\approx0.109|a_{1g}\uparrow\rangle-
  0.061i\left(|e_{g1}^{\prime}\downarrow\rangle+
    i|e_{g2}^{\prime}\downarrow\rangle\right)-
  0.700\left(|e_{g1}\downarrow\rangle+i|e_{g2}\downarrow\rangle\right).
 \nonumber
\end{align}
Due to the time-reversal symmetry, all double group pairs are satisfied 
with $\mathcal{T}|\tau_a\tilde{\uparrow}\rangle=e^{i\delta_a}|
\tau_a\tilde{\downarrow}\rangle$, where $\mathcal{T}$ is the 
time-reversal operator and $\delta_a$ is the phase term raised by
numerical process. However, $\delta_a$ does not cause any calculation
complexity.

\section{Hopping Hamiltonian}

To describe the hopping interaction, we adopted the tight binding method based on 
the linear combination of atomic obritals (LCAO). 
Because the double group state $\tau$ with $\tilde{\sigma}$ isospin 
is expressed by $|\tau\tilde{\sigma}\rangle = 
  \sum_{\mu\sigma} U_{\tau\tilde{\sigma},\mu\sigma} |\mu\sigma\rangle$,
where $ U_{\tau\tilde{\sigma},\mu\sigma}$ is the unitary transformation
between $\tau\tilde{\sigma}$ and conventional $d$ orbital and spin,
the $pd$-hopping matrix is written by
$t_{pd}(\tau\tilde{\sigma};p\sigma)=\sum_{\mu\sigma^{\prime}}
 U_{\tau\tilde{\sigma},\mu\sigma^{\prime}} \delta_{\sigma\sigma^{\prime}}
 \langle p|V_h|\mu\rangle$, where $\langle p|V_h|\mu\rangle$ is the
hopping strength between $p$ and $d$ atomic orbitals.
$\langle p|V_h|\mu\rangle$ is a function of 
two parameters ($t_{pd\sigma}$ and $t_{pd\pi}$) and 
normal displacement vector between Ir and O.
Next, we estimated the effective hopping between nearest neighboring (NN) Ir's 
based on the second-order perturbation because the charge transfer energy ($\Delta$)
is much larger than the $pd$-hopping strengths.
In this limit,
$t_{dd}(\tau\tilde{\sigma};\tau^{\prime}\tilde{\sigma}^{\prime})$
between NN double group states $\tau\tilde{\sigma}$
and $\tau^{\prime}\tilde{\sigma}^{\prime}$
is calculated by summing 
$\sum_{p\sigma}\frac{t_{pd}(\tau\tilde{\sigma};p\sigma)
 t_{pd}^*(\tau^{\prime}\tilde{\sigma}^{\prime};p\sigma)}
{\sqrt{(\Delta+\epsilon_{\tau})(\Delta+\epsilon_{\tau^{\prime}})}}$
for two different Ir-O-Ir paths.
The hopping Hamiltonian between $i$ and $j$-th Ir's is as following:
\begin{equation}
H_{ij}=\sum_{\tau_i\tilde{\sigma}_i\tau^{\prime}_j\tilde{\sigma}^{\prime}_j}
   t_{dd}(\tau_i\tilde{\sigma}_i;\tau^{\prime}_j\tilde{\sigma}^{\prime}_j)
   c_{\tau^{\prime}_j\tilde{\sigma}^{\prime}_j}^{\dagger}
   c_{\tau_i\tilde{\sigma}_i}+\textit{h.c.}.
\end{equation}
Note that the full hopping matrix ($10\times10$) is Hermitian.
Namely, the full hopping matrix elements satisfy the relation such that 
$\langle \tau_a\tilde{\downarrow} | H_t |\tau_b\tilde{\uparrow} \rangle
= \overline{ \langle \tau_b\tilde{\uparrow} | H_t 
| \tau_a\tilde{\downarrow}\rangle}$.
Table~\ref{tb1} and~\ref{tb2} present the hopping matrix between neighboring
Ir's along the $y$-axis.
The hopping between $\tau_1\tilde{\downarrow}$ and $\tau_3\tilde{\uparrow}$ 
($0.2074$) is strongest one, which is
about ten times larger than that between $\tau_3\tilde{\uparrow}$'s ($0.0219$).

\begin{table}[!h]
\caption {
Hopping parameters between neighboring double group state with same isospins 
in units of eV.
}
\label{tb1}
\begin{ruledtabular}
\begin{tabular}{c | c c c c c}
 & $\tau_1\tilde{\uparrow}$ & $\tau_2\tilde{\uparrow}$ & $\tau_3\tilde{\uparrow}$ &
 $\tau_4\tilde{\uparrow}$ & $\tau_5\tilde{\uparrow}$ \\
\hline
 $\tau_1\tilde{\uparrow}$ &  0.1362    &  0.0797$i$ & 0.0213$i$ &  0.1348$i$ &
-0.1125$i$ \\
 $\tau_2\tilde{\uparrow}$ & -0.0797$i$ & -0.0520    & -0.0975   &  0.1306    &
-0.1186    \\
 $\tau_3\tilde{\uparrow}$ & -0.0213$i$ & -0.0975    &  0.0219   & -0.1301    &
 0.1119    \\
 $\tau_4\tilde{\uparrow}$ & -0.1348$i$ &  0.1306    & -0.1301   &  0.0558    & 
 0.0709    \\
 $\tau_5\tilde{\uparrow}$ &  0.1125$i$ & -0.1186    &  0.1119   &  0.0709    &
 0.0976    \\
\end{tabular}
\end{ruledtabular}
\end{table}

\begin{table}[!h]
\caption {
Hopping parameters between neighboring double group state with different isospins
in units of eV.
}
\label{tb2}
\begin{ruledtabular}
\begin{tabular}{c | c c c c c}
 & $\tau_1\tilde{\uparrow}$ & $\tau_2\tilde{\uparrow}$ & $\tau_3\tilde{\uparrow}$ &
 $\tau_4\tilde{\uparrow}$ & $\tau_5\tilde{\uparrow}$ \\
\hline
 $\tau_1\tilde{\downarrow}$ &  0.0000    & -0.0212    & -0.2074    & -0.0933    &  
 0.0978    \\
 $\tau_2\tilde{\downarrow}$ &  0.0212$i$ &  0.0000    & -0.1504    &  0.1426    & 
-0.1322    \\
 $\tau_3\tilde{\downarrow}$ &  0.2074    & -0.1504$i$ &  0.0000    &  0.0362$i$ &  
-0.0679$i$ \\
 $\tau_4\tilde{\downarrow}$ &  0.0933$i$ & -0.1426    &  0.0362    &  0.0000    & 
-0.0130    \\
 $\tau_5\tilde{\downarrow}$ &  0.0978    &  0.1322$i$ & -0.0679$i$ &  0.0130$i$ &  
 0.0000    \\
\end{tabular}
\end{ruledtabular}
\end{table}


\begin{thebibliography}{99}
\bibitem{Octa} A monoclinic Na$_2$IrO$_3$ ($C2/m$ space group) 
   suffers from several structural deformations such as orthorhombic 
   distortion, octahedra rotations, and trigonal distortion~\cite{Choi}; the
   latter has the most impact on the electronic structure~\cite{Foyevtsova}.
\bibitem{Hspace}
  The Hilbert space is classified into seven subspaces:
  $\mathbf{H}_1$ of four $\bar{D}$'s of $d^5$,
  $\mathbf{H}_2$ of one or more $\bar{Q}$ among four $d^5$ configurations,
  $\mathbf{H}_3$ of two $\bar{D}$'s of $d^5$ and 
  an $\bar{A}$-$\bar{S}$ $e$-$h$ pair,
  $\mathbf{H}_4$ of two $\bar{D}$'s of $d^5$ and 
  an $\bar{A}$-$\bar{T}$ $e$-$h$ pair,
  $\mathbf{H}_5$ of two $\bar{D}$'s of $d^5$ and 
  an $\bar{A}$-$\bar{P}$ $e$-$h$ pair ,
  $\mathbf{H}_6$ of two $\bar{D}$'s of $d^5$ and 
  an $\bar{A}$-$\bar{P}^{\prime}$ $e$-$h$ pair, 
  and others ($\mathbf{H}_7$).
\bibitem{Hop} In our calculation, the next or third NN hopping and the 
direct Ir-Ir hopping are not included, because the strength of 
the NN hopping mediated by oxygen is much larger than other 
hoppings~\cite{Mazin1}.
\bibitem{Rixs} 
   To calculate the RIXS spectra,
   we assumed that both incident and outgoing angles of x-ray
   are $45^{\circ}$, and the incident and outgoing beams have $\sigma$
   and arbitrary polarizations, respectively.
\end{thebibliography}
\end{document}